\documentclass{SciPost}

\hypersetup{
    colorlinks,
    linkcolor={red!50!black},
    citecolor={blue!50!black},
    urlcolor={blue!80!black}
}

\usepackage[bitstream-charter]{mathdesign}
\DeclareSymbolFont{usualmathcal}{OMS}{cmsy}{m}{n}
\DeclareSymbolFontAlphabet{\mathcal}{usualmathcal}

\fancypagestyle{SPstyle}{
\fancyhf{}
\lhead{\colorbox{scipostblue}{\bf \color{white} ~SciPost Physics }}
\rhead{{\bf \color{scipostdeepblue} ~Submission }}

\fancyfoot[C]{\textbf{\thepage}}
}

\begin{document}

\pagestyle{SPstyle}

\begin{center}{\Large \textbf{\color{scipostdeepblue}{
Emergent Pair Density Wave Order Across a Lifshitz Transition\\
}}}\end{center}

\begin{center}\textbf{
Luhang Yang\textsuperscript{1$\star$},
Elbio Dagotto\textsuperscript{1, 2} and
Adrian E. Feiguin\textsuperscript{3}
}\end{center}

\begin{center}
{\bf 1} Department of Physics and Astronomy, University of Tennessee, Knoxville, Tennessee 37996, USA
\\
{\bf 2} Materials Science and Technology Division, Oak Ridge National Laboratory, Oak Ridge, Tennessee 37831, USA
\\
{\bf 3} Department of Physics, Northeastern University, Boston, Massachusetts 02115, USA
\\[\baselineskip]
$\star$ \href{mailto:email1}{\small luhangyang@utk.edu}\
\end{center}

\section*{\color{scipostdeepblue}{Abstract}}
\textbf{\boldmath{%
We numerically investigate the telltale signs of pair-density-wave order (PDW) in the Kondo-Heisenberg chain by focusing on the momentum resolved spectrum in different parameter regimes. Density matrix renormalization group calculations reveal that this phase is characterized by a dispersion with two minima and four Fermi points, indicating the emergence of an effective next-nearest-neighbor hopping that arises as a second-order effect to avoid magnetic frustration. The pairs appear in the spectrum as in-gap bound states with weight concentrated in the hole pockets. The low-energy physics can be understood by means of a generalized $t-J$ model with next-nearest-neighbor hopping. 
Our results offer a guide for searching for experimental signatures, and for other models that can realize PDW physics.
}}

\vspace{\baselineskip}

\noindent\textcolor{white!90!black}{%
\fbox{\parbox{0.975\linewidth}{%
\textcolor{white!40!black}{\begin{tabular}{lr}%
  \begin{minipage}{0.6\textwidth}%
    {\small Copyright attribution to authors. \newline
    This work is a submission to SciPost Physics. \newline
    License information to appear upon publication. \newline
    Publication information to appear upon publication.}
  \end{minipage} & \begin{minipage}{0.4\textwidth}
    {\small Received Date \newline Accepted Date \newline Published Date}%
  \end{minipage}
\end{tabular}}
}}
}


\vspace{10pt}
\noindent\rule{\textwidth}{1pt}
\tableofcontents
\noindent\rule{\textwidth}{1pt}
\vspace{10pt}


\section{Introduction}
\label{sec:intro}
The pair density wave state (PDW) was proposed as a possible scenario to interpret the anomalous superconducting phase in the La$_{1.875}$Ba$_{0.125}$CuO$_4$ \cite{cuprate_LBCO,LBCO2}, where pairing is intertwined with a stripe or charge density wave (CDW) order. This stripe/PDW phase appears in the underdoped regime where the spin gap opens. Similar PDW order was also suggested as a candidate phase in Bi2212 \cite{cuprate_BSCCO,BSCCO2}, Bi2201 \cite{Pb-Bi2201}, Kagome superconductors \cite{chen2021roton}, iron-based superconductors \cite{liu2023pair}, and transition metal dichalcogenides \cite{Venderley2019}.

In the PDW phase, the superconducting order parameter exhibits spatial oscillations, and electron pairs acquire a finite center-of-mass momentum.
This novel pairing order is naturally distinct from the Fulde–Ferrell–Larkin–Ovchinnikov (FFLO) state \cite{FF_state,LO_state}, which is induced by an external magnetic field and an imbalance between majority and minority spin populations. It also differs from the uniform superconducting (SC) state, where the spatial average of the order parameter is non-vanishing. 

Many numerical and theoretical studies have been conducted to solve the puzzles surrounding the PDW state, and its origin still remains unclear \cite{Chubukov2015,PLee2014, Fradkin_review,agterberg2020physics,berg2009striped,agterberg2008dislocations,shousheng2007,Berg2010,hongchen2023,hongchen2,my_3band,Jaefari2012,Soto2015,Chun2018,Hassanieh2009}. 
In a seminal work, a computational and theoretical study of the Kondo-Heisenberg (KH) model \cite{Berg2010} demonstrated that the PDW state coexists with charge density wave (CDW) order in the spin gapped phase.
Although conducted in one spatial dimension and on a model not obviously related to high-Tc superconductivity, this study provides valuable insights on a numerically solvable Hamiltonian and has motivated further research in this new direction \cite{Fangze2024}. 

The opening of the spin gap has long been believed to be a precursor to the emergence of superconductivity, PDW, and CDW instabilities \cite{Berg2010,Jaefari2012,Zachar2001,Zachar2001_2}. However, the exact mechanism driving the PDW transition is not well understood.
A recent theoretical mean-field study of superconductivity on the triangular lattice \cite{Jin2022} pointed out that the Fermi surface geometry can impact the PDW formation. 
Despite this progress, our understanding of the interplay between antiferromagnetism, Fermi surface geometry, and PDW order is still far from being settled. 

In this work, we revisit the one-dimensional Kondo-Heisenberg model by paying particular attention to the topology of the Fermi surface and the excitation spectrum.
Our results show that the emergence of the PDW is closely related to a ``hump'' feature in the electronic momentum distribution function (MDF), which can be interpreted as the manifestation of a Lifshitz transition \cite{liu2010evidencelifshitz,hinlopen2024lifshitz} from a Fermi surface with two momenta $\pm k_F$, to one with four. We attribute this transition to an effective next-nearest-neighbor hopping \cite{Nocera2020} arising from the magnetic interaction with the localized spins.

\section{Model and Method}
We investigate the Kondo-Heisenberg chain, described by the following Hamiltonian:

\begin{eqnarray}
   \nonumber H_{KH} &=& -t\sum_{i\sigma}(c_{i,\sigma}^\dagger c_{i+1,\sigma} + h.c.) + J_H \sum_{i}\vec{S}_i\cdot \vec{S}_{i+1} \\ 
    &+& J_K \sum_{i}\vec{s}_i\cdot \vec{S}_{i},
\label{KH}
\end{eqnarray}
where $c^\dagger_{i\sigma}$ creates an electron of spin $\sigma$ on the
$i^{\rm th}$ site along a chain of length $L$. The localized spin $\vec{S}_i$ interact with each other via an exchange term, and with the conduction electrons through a Kondo term, parametrized by $J_H$ and $J_K$, respectively. Both interactions are positive (antiferromagnetic), as usually assumed for heavy fermions systems \cite{Tsunetsugu_RMP}. We take the inter-atomic distance as unity and we shall express all energies in units of the hopping parameter $t$. 

The phase diagram of the Kondo-Heisenberg model with hole doping was first determined in Ref.\cite{Sikkema1997}. At 1/8 hole doping and $J_H\lesssim 2$, this model exhibits a transition from a spin-gapped phase to a Luttinger Liquid phase with increasing $J_K$ \cite{Sikkema1997, fermi_surface_1,fermi_surface_2}.  
The spin gap has non-monotonic behavior, growing exponentially immediately after turning on the Kondo coupling $J_K$ \cite{Sikkema1997, Fujimoto1994,fermi_surface_1,fermi_surface_2}, and decreasing to zero for large enough $J_K$, when it becomes a Luttinger liquid (LL). When $J_K>>J_H$, the electrons become strongly coupled to local impurities to form heavy fermions with a large Fermi surface. 
In this regime, one can think of the half-filled chain as a vacuum of localized singlets. By doping it, each hole partners with a dangling spin on the Heisenberg chain, playing the role of fermionic quasiparticles interacting via the Heisenberg exchange $J_H$. As a result, the physics corresponds to that of a a modified $t-J$ chain with density $|1-n|$ where the holes play the role of the electrons \cite{Sikkema1997}.
Between these two limits, at intermediate values of $J_K$, the system displays strong PWD and CDW tendencies.

In the following, we choose a parameter regime that allows us to study the transition to a PDW phase. Specifically, 
we present results for $J_K$ varying between 1 and 4 and $J_H$ between 1 and 2, within the spin-gapped (we present results for the spin gap in Appendix \ref{sec:app_spin_gap}) phase for the 1/8 hole doped Kondo-Heisenberg model \cite{Sikkema1997}. This parameter range has also been shown to display a binding tendency near half-filling \cite{Xavier}, and is smoothly connected to small values of $J_K$ and $J_H$, that are more realistic representations of materials. \cite{Sikkema1997,Xavier}

We use the density matrix renormalization group
(DMRG) method \cite{White1992,white1993} to study chains of lengths from 32 to 80 sites, using a bond dimension up to $m=1600$ to ensure a truncation error smaller than $10^{-6}$. We also use time-dependent DMRG (tDMRG) \cite{white2004,daley2004,vietri,Paeckel2019} to compute the photoemission and inverse-photoemission spectra. The tDMRG is implemented by using a Suzuki-Trotter decomposition of the time-evolution operator. In order to get the momentum and frequency resolved spectra, we calculate the single particle correlation functions in space and time $\langle c^\dagger_i(T)c_j(0)\rangle$ and $\langle c_i(T)c^\dagger_j(0)\rangle$ in time steps of $\tau=0.05$ to times up to $T=60$, and then Fourier transform the results to momentum and frequency with an exponential envelope that yields a Lorentzian lineshape of width $\epsilon=0.1$. While the truncation error grows to $10^{-6}$ at longer times, the effects are mitigated by the exponential envelope, which also helps to alleviate ringing artifacts arising from Fourier transforming in a finite time window.

\section{Results}
\label{sec:another}
\begin{figure}
	\centering
  \includegraphics[width=0.7\textwidth]{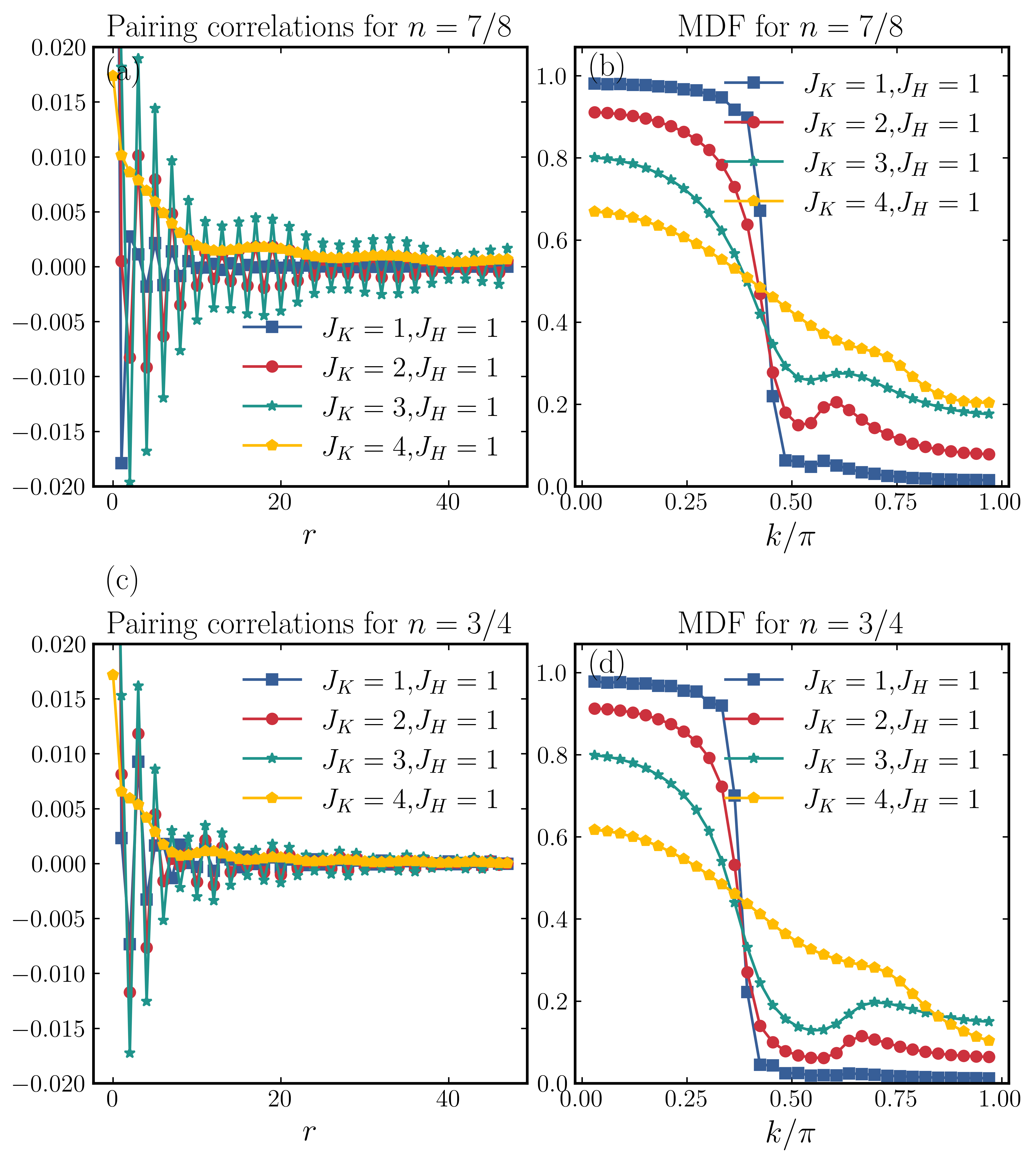} 	
  \caption{(a) Pairing correlations as a function of distance from the reference site for $L=64$ and $N=56$; (b) momentum distribution function for $L=32$ and $N=28$ (c) Pairing correlations for $L=64$ and $N=48$, (d) momentum distribution function for $L=32$ and $N=24$. } \label{fig:pdw}
\end{figure}

To understand the connections between Kondo physics, antiferromagnetism and the pairing mechanism, 
we probe the pair-pair correlations $P_s(i,j) = \langle \Delta^\dagger_i\Delta_j\rangle$, with
\begin{equation}
\Delta^\dagger_i = \frac{1}{\sqrt{2}}(c^\dagger_{i,\downarrow}c^\dagger_{i+1,\uparrow} - c^\dagger_{i,\uparrow}c^\dagger_{i+1,\downarrow}).
\label{Psr}
\end{equation}
We notice that, unlike the doublon pairs in the negative-$U$ Hubbard model \cite{Feiguin2009}, pairs here are single electrons forming a singlet on nearest-neighbor sites.
By fixing $J_H=1$ and increasing $J_K$, the PDW order in the KH model is first enhanced, then transitions to a uniform superconducting order (Fig.\ref{fig:pdw} (c)). In Fig.\ref{fig:pdw} (a) we show the pairing correlations for KH model (Eq.\ref{KH}) 
in a linear scale. At $1/8$ hole doping and $J_K=3$, $J_H=1$, the PDW is the dominant order and is accompanied by a subsidiary charge density wave (CDW); when $J_K=4$, $J_H=1$, the KH model exhibits quasi-long-ranged uniform superconducting order with no oscillatory component. 
We show he results for spin-spin, density-density, single-particle and pairing correlations across the transition in Appendix \ref{sec:app_corr}, where the emergence of the PDW phase and the transition from PDW to uniform SC can be seen by comparing these correlations.

When the Kondo coupling is in the intermediate regime ( {\it i.e.} of the order of the bandwidth), $J_K=1$ to 4, a sizable spin gap opens, with the onset of short-range antiferromagnetic correlations, charge order, and a pairing instability characterized by an order parameter that oscillates in space with momentum vector $K_{PDW}=\pi$.

To further characterize the origin of the PDW phase, we present results for the momentum distribution function (MDF), obtained by Fourier transforming the single particle correlations $G(i,j) = \langle c^\dagger_{i,\uparrow} c_{j,\uparrow}\rangle$. Results 
for $J_K=1,2,3,4$ are shown in Fig.\ref{fig:pdw}. As $J_K$ increases, we observe the appearance of a ``hump'' feature, that evolves to a singularity with a larger Fermi surface, corresponding to a transfer of spectral weight from low to high momenta. This excess of spectral weight spans the range between $k_{F2}=\pi-k_{F1}$ and $k=\pi$, where $k_{F1}\simeq k_F=\pi/2n$. This feature in the momentum distribution has previously been seen in similar models \cite{fermi_surface_1,fermi_surface_2}. 
As expected for interacting one-dimensional systems, the Fermi points at $k_{F2}$ are not defined by sharp discontinuities in the MDF. Thus, the four Fermi points (including $k_{F1}$ and $k_{F2}$) in our discussions hereafter are identified via the spectral weight at the Fermi level in $A(k,\omega)$.

\begin{figure*}
	\centering
   \includegraphics[width=1.0\textwidth]{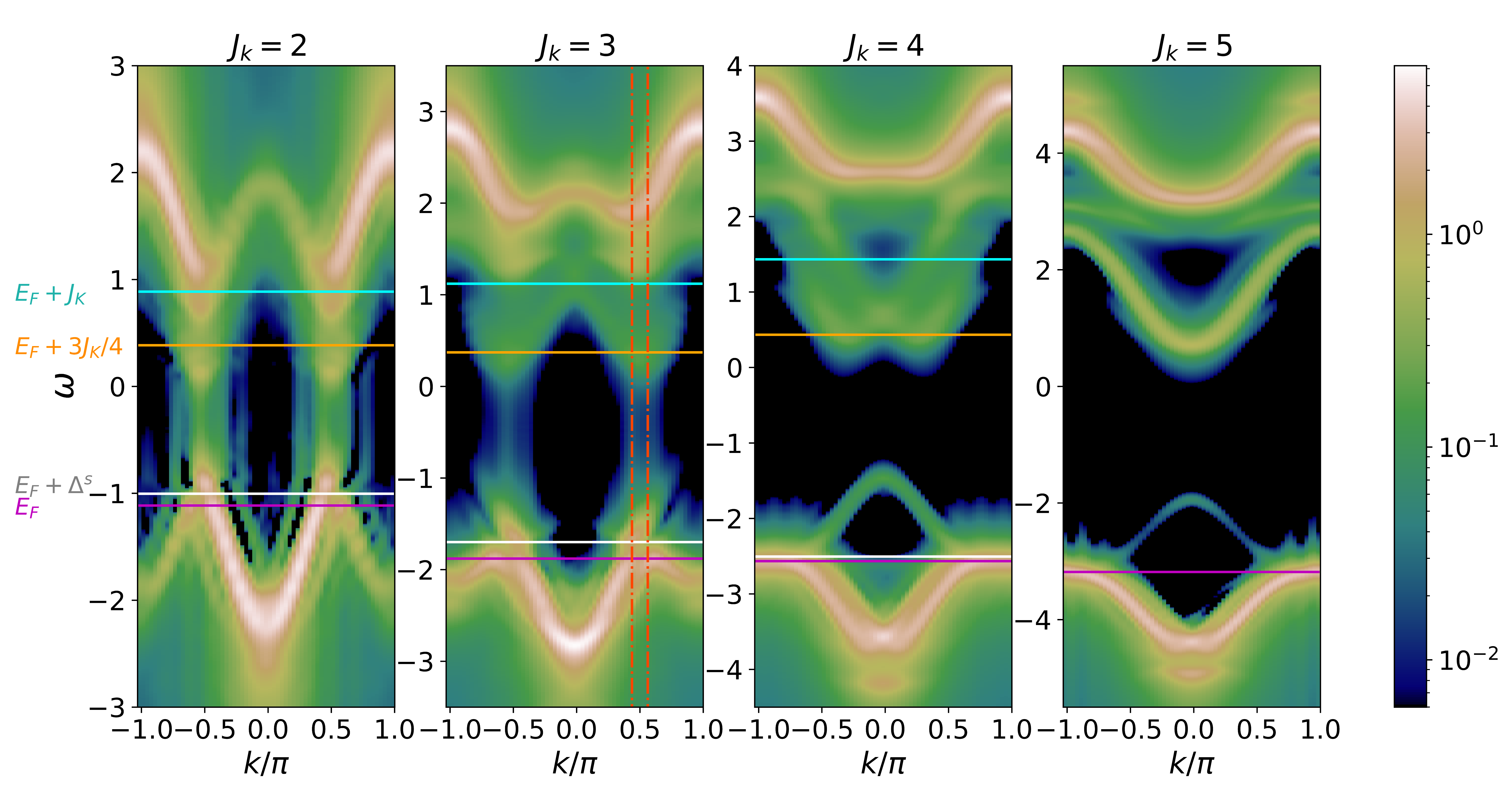}
	\caption{Photoemission and inverse-photoemission spectra for $J_H=1$, $J_K=2$ to 5 for the Kondo-Heisenberg model. The single-particle removal spectra and single-particle addition spectra are separated by the solid magenta lines noting the Fermi levels with $1/8$ hole doping. The solid white lines mark the spin (and charge) gaps; the solid orange lines represent the energy of $3 J_K/4$ above Fermi level; and the solid blue lines are at $J_K$ above Fermi level. The vertical lines demark $k_F$ ($k_{F1}$)} and $\pi-k_F$ \textcolor{blue}{($k_{F2}$)} respectively. The figures are plotted in log-scale colormaps to show the faint features. Same color scale is applied in all panels. \label{fig:photoemission}
\end{figure*}

To elucidate the origin of this feature, we compute the photoemission spectra for the KH model with $J_H=1$, and $J_K$ ranging from 2 to 5, shown in Fig.\ref{fig:photoemission}. We use a log scale for the spectral weight to visualize faint features in the spectrum (notice that this choice tends also to accentuate small ringing errors due to the Fourier transform). These four cases represent three scenarios: $J_K=2$ and $J_K=3$ correspond to the emergence and enhancement of the PDW order; the case of $J_K=4$ lies at the onset of uniform SC and the fading of the PDW; $J_K=5$ depicts the formation of coherent heavy fermions.
Furthermore, the results for $J_K=3$ and $J_K=4$ also illustrate the cases with and without the ``hump'' in the MDF, respectively.

As $J_K$ increases, we observe the vanishing of the in-gap states caused by the Kondo interaction. There are two high energy features: a more intense, high energy band at $\omega \sim J_K$, corresponding to a singlet-triplet excitation, and a fainter one at energy $\sim 3J_K/4$ due to doublon excitations, that are actually the edge of a continuum. More interestingly, we also observe in-gap excitations appearing right above the Fermi level (see Fig.\ref{fig:photoemission_1} in Appendix \ref{sec:app_small_large_JK} for photoemission results at $J_K=1$). These in-gap states, correspond to an added particle on top of the condensate and are separated from the valence band by the spin gap \cite{Feiguin2009}. 
The most dramatic effect is the development of two minima and four Fermi points in the lower band, below the Fermi level. While these features are not clear due to the insufficient momentum resolution, they are more evident at density $n=5/8$ (See SM).  
This change in the topology of the Fermi surface can only be attributed to an effective next-nearest-neighbor hopping arising from a competition between the magnetic order and the Kondo coupling $t_2\sim t^2/J_K$: Since the Heisenberg chain has short-range antiferromagnetic correlations, an electron with a given spin projection has to hop two sites in order to avoid a ferromagnetic alignment with a localized spin.
At $J_K=4$, the PDW becomes a uniform SC order, the two local minima are replaced by a flat band region centered at $k=\pi$, and the in-gap states at low energy are replaced by a wide gapless band with minima at $\sim 2k_F$. 
At $J_K\sim 5$, as the system transitions from a SC to a LL phase, the spectral weight becomes more uniformly distributed in momentum, and the dispersion recovers its more familiar cosine-like profile with only an effective nearest-neighbor hopping $t_{eff}\sim 0.5$, as expected in the heavy-fermion phase \cite{Tsunetsugu_RMP,spectrum_kl}.

\begin{figure}
	\centering
  \includegraphics[width=0.7\textwidth]{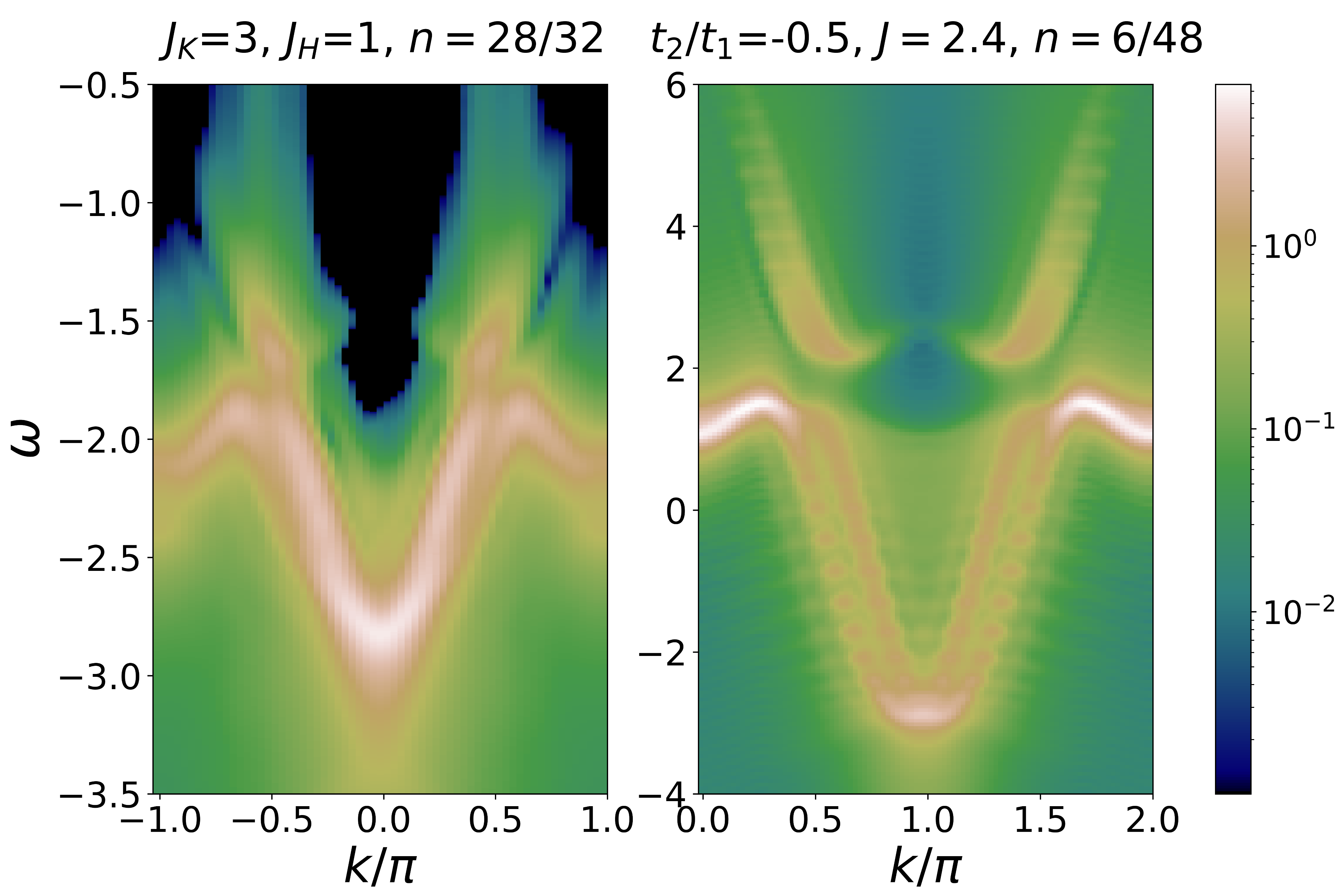} 	
  \caption{Left: Single-particle removal spectrum for the Kondo-Heisenberg model; Right: Same for the $t_1-t_2-J$ model {\it in the hole language} (see text). The spectral weight is in log-scale. Same color scale is applied
to both panels.} \label{fig:tJ}
\end{figure}

To further understand the low energy physics in the PDW phase, we investigate the connection between the KH and the $t_1-t_2-J$ model \cite{myt1_t2_J}:
\begin{eqnarray}
    H_{t_1-t_2-J} & = & -
    \sum_{i,\sigma}t_{1}(c_{i,\sigma}^\dagger c_{i+1,\sigma} + h.c.) \\ \nonumber
    & - & \sum_{i,\sigma}t_{2}(c_{i,\sigma}^\dagger c_{i+2,\sigma} + h.c.)  \\
    \nonumber
    & + & J\sum_{i}(\vec{S}_i\cdot \vec{S}_{i+1} - 
\cfrac{1}{4} n_in_{i+1}), 
\label{hami_tj}
\end{eqnarray}
where $n_{i}=\sum_\sigma c^\dagger_{i\sigma}c_{i\sigma}$ is the electron number operator, and we implicitly apply the constraint forbidding double occupancy. We compare the spectral functions of the KH and the $t_1-t_2-J$  models at densities $n=7/8$ and $1/8$, respectively. In Fig.\ref{fig:tJ} we show results for the KH model with $J_H=1$,$J_K=3$, and the $t_1-t_2-J$ model with $t_1=1$; $t_2=-0.5$ and $J=2.4$, deep in the PDW phase. In the large $J_K/J_H$ limit, electrons in the KH model become equivalent to holes in the $t_1-t_2-J$ model. To compare the latter to the former, one needs to apply a particle-hole transformation and look at its hole spectral function (related to the electronic one by a sign change in $\omega$ and a $\pi$-shift in momentum). 
Both spectra show identical features, which is a strong indication that the low energy physics of both models is similar in this regime.

Since the dispersion with two local minima is a feature that emerges at half filling --see Appendix \ref{sec:app_filling} for additional photoemission results--, and persists upon doping (Fig.\ref{fig:photoemission}), we expect that the PDW phase will be stable in a wide range of densities on the hole doped side. To examine this, we computed the ground state correlations at different densities. In  Fig.\ref{fig:pdw} (c) we present results at filling factor $n=0.75$, illustrating a persisting PDW instability. In addition, the MDF (Fig.\ref{fig:pdw} (d)) also shows a similar transition as the lightly doped case.

\section{Conclusion}
The PDW phase in the KH model is mediated by the interplay between magnetic order and Kondo exchange and emerges at the same time as the electronic band undergoes a Lifshitz transition from a ``Fermi surface'' with two Fermi points, to one with four, and a double minimum. The role of magnetism is two-fold: (i) it induces an effective next-nearest-neighbor hopping that affects the topology of the electronic band and (ii) acts as the pairing glue \cite{Scalapino1986,Maier2007,Maier2008,Kyung2009,Khatami2009}. 

In the intermediate $J_K$ regime --$J_K$ of the order of the bandwidth-- where the PDW is stabilized, we observe that the band acquires two minima and four Fermi points, indicating the emergence of an effective next-nearest-neighbor hopping $t_2$. This second order process allows electrons to gain kinetic energy without paying magnetic frustration. We find that the low energy physics in the PDW phase is in correspondence with the results of the $t_1-t_2-J$ model in the dilute regime, where a PDW phase emerges at large hole doping and intermediate values of $J$ \cite{myt1_t2_J}. The connection between the KH model and the $t-J$ model in the strong $J_K$ regime was established in Ref.\cite{Sikkema1997} but the need for an additional next-nearest-neighbor hopping was overlooked.

In the ground state, the momentum distribution develops a ``hump'' feature, with a ``trench'' carved between momenta $k_{F1}$ and $k_{F2}$, and the pair-pair correlations display a modulation with momentum $K_{PDW}=\pi$. 
It is useful to contrast these observations to the FFLO case, where the four Fermi momenta correspond to those of the majority and minority spin, $\pm k_{F\uparrow},\pm k_{F\downarrow}$, and the superconducting order parameter oscillates with phase $K_{FFLO}=k_{F\uparrow}-k_{F\downarrow}$. 
Hence, by analogy, a naive expectation would be to assume that the PDW order would oscillate with phase $K_{PDW}=k_{F1}$-$k_{F2}$. 
Instead, the center-of-mass momentum of the pairs is fixed at $K_{PDW}=\pi=k_{F1}+k_{F2}$, regardless of the hole density. 
This implies that pairs are formed by electrons at momenta $k_{F1}$ and $k_{F2}$, instead of $k_{F1}$ and $-k_{F2}$, as already observed in Refs.\cite{Zachar2001,Zachar2001_2,Berg2010} (see also Refs.\cite{agterberg2020physics,Venderley2019}). 
In fact the two modes at $k_{F1} + k_{F2} = \pi$ and $k_{F1} - k_{F2} = 2k_{F1} + \pi$ are also found in Ref. \cite{Berg2010}, where it is revealed that the excitation at $2k_{F1} + \pi$ is a ``charge zero'' mode, whilst the one at $\pi$ carries charge 2$e$. Since the Cooper pair possesses charge 2$e$ instead of charge zero, the momentum vector of the PDW is locked at $k_{F1} + k_{F2}$. This reflects the ``intertwined'' nature of the PDW, which contains contributions from both the localized spins and the charge density wave.

Our study of the ground state properties and the excitation spectrum of the KH model demonstrates that the emergence of the PDW is associated to a strong renormalization of the band with a change in the topology of the Fermi surface due to an effective next-nearest-neighbor hopping. In models that already explicitly include this additional next-neighbor hopping term, the PDW is already manifested, indicating that it may be a fundamental ingredient for its realization.
Since the $t-J$ model, the three-band Hubbard model, and the two-leg ladder model proposed in Ref.\cite{Jaefari2012} are all connected to the KH model \cite{Berg2010,hongchen2023,my_3band,Jaefari2012}, it is plausible to postulate that the mechanism we propose applies to all of these models.
The identification of the Lifshitz transition in the intermediate interacting regime bridges a gap between the established theoretical analysis and the more physically relevant and experimentally accessible parameter regime, offering a path toward experimental confirmation in materials such as organic radical chains \cite{li2024kondochains}.

\section*{Acknowledgements}
LY and ED are supported by the U.S. Department of Energy (DOE), Office of Science, Basic Energy Sciences (BES), Materials Sciences and Engineering Division. AEF is supported by the National Science Foundation under grant
No. DMR-1807814. This manuscript has been authored by UT-Battelle, LLC, under contract DE-AC05-00OR22725 with the US Department of Energy (DOE). The US government retains, and the publisher, by accepting the article for publication, acknowledges that the US government retains a nonexclusive, paid-up, irrevocable, worldwide license to publish or reproduce the published form of this manuscript, or allow others to do so, for US government
purposes. DOE will provide public access to these results of federally sponsored research in accordance with the DOE Public Access Plan (https://www.energy.gov/DOE Public Access Plan).

\begin{appendix}
\numberwithin{equation}{section}

\section{Comparison of correlations for different $J_K$}
\label{sec:app_corr}

\begin{figure}
	\centering
  \includegraphics[width=1.0\textwidth]{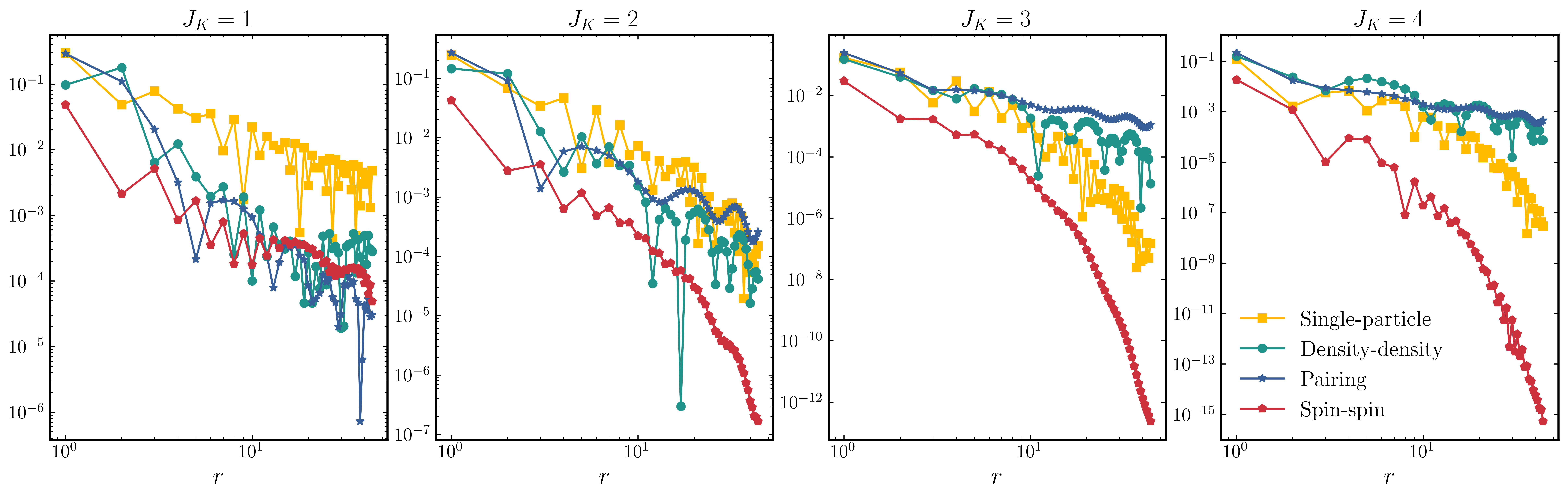}
	\caption{Spin-spin, density-density, single-particle, and pairing correlations across the transition. The results are obtained using DMRG for chains of length $L=64$. The correlations are plotted in a log-log scale. } \label{fig:correlations}
\end{figure}

We calculated the spin-spin, density-density, single-particle, and pairing correlations across the transition as $J_K$ increases. 
The spin-spin correlations are given by:
\begin{equation}
S(r) = \langle S^z_0S^z_r\rangle;
\label{Sr}
\end{equation}
the density-density correlations are defined as: 
\begin{equation}
D(r) = \langle n_0n_r\rangle - \langle n_0\rangle \langle n_r\rangle;
\label{NNr}
\end{equation}
the single particle correlations are defined as: 
\begin{equation}
G(r) = \langle c^\dagger_{0,\uparrow} c_{r,\uparrow}\rangle;
\label{Gr}
\end{equation}
and the pairing correlations are given in Eq.\ref{Psr} in the main text.
The values of the local spin modulation $\langle S^z_i\rangle$ are numerically negligible for each site $i$; thus, the spin-spin correlations behave in the same way as $\langle S^z_0S^z_r\rangle - \langle S^z_0\rangle\langle S^z_r\rangle$. The local density $\langle n_i\rangle$ modulates around its average value at the corresponding filling density; thus, we use Eq.\ref{NNr} to represent the density-density correlations.
To minimize the oscillations induced by the open boundaries, we exclude 10 sites at each end and show the results in the central region. We take the 11th site as the reference site, and the distance $r$ is defined as the distance from this reference site. The correlations are plotted in a log-log scale to show the power-law decaying behavior of the quasi-long range order. 

When $J_K$ is small ($J_K=1$), all correlations exhibit power-law decay; thus, the curves in the log-log scale plot are linear. As $J_k$ increases, the pairing order is enhanced and spin order decays faster. When the system enters the PDW phase at $J_K=3$, the pairing correlations become dominant, and are accompanied by a sub-dominant charge order. Meanwhile, the spin-spin correlation decays exponentially, indicating the opening of a spin gap, as we will show in the next section. The pairing correlations in Fig.\ref{fig:pdw}
of the main text show oscillations around zero in this parameter regime; thus, the behavior is consistent with a pair density wave. By further increasing $J_K$ to $J_K=4$, these oscillations around zero eventually disappear, and the pairing order evolves into a uniform SC order. 

\section{Spin gaps and spin-spin correlations}
\label{sec:app_spin_gap}

\begin{figure}
	\centering
  \includegraphics[width=0.55\textwidth]{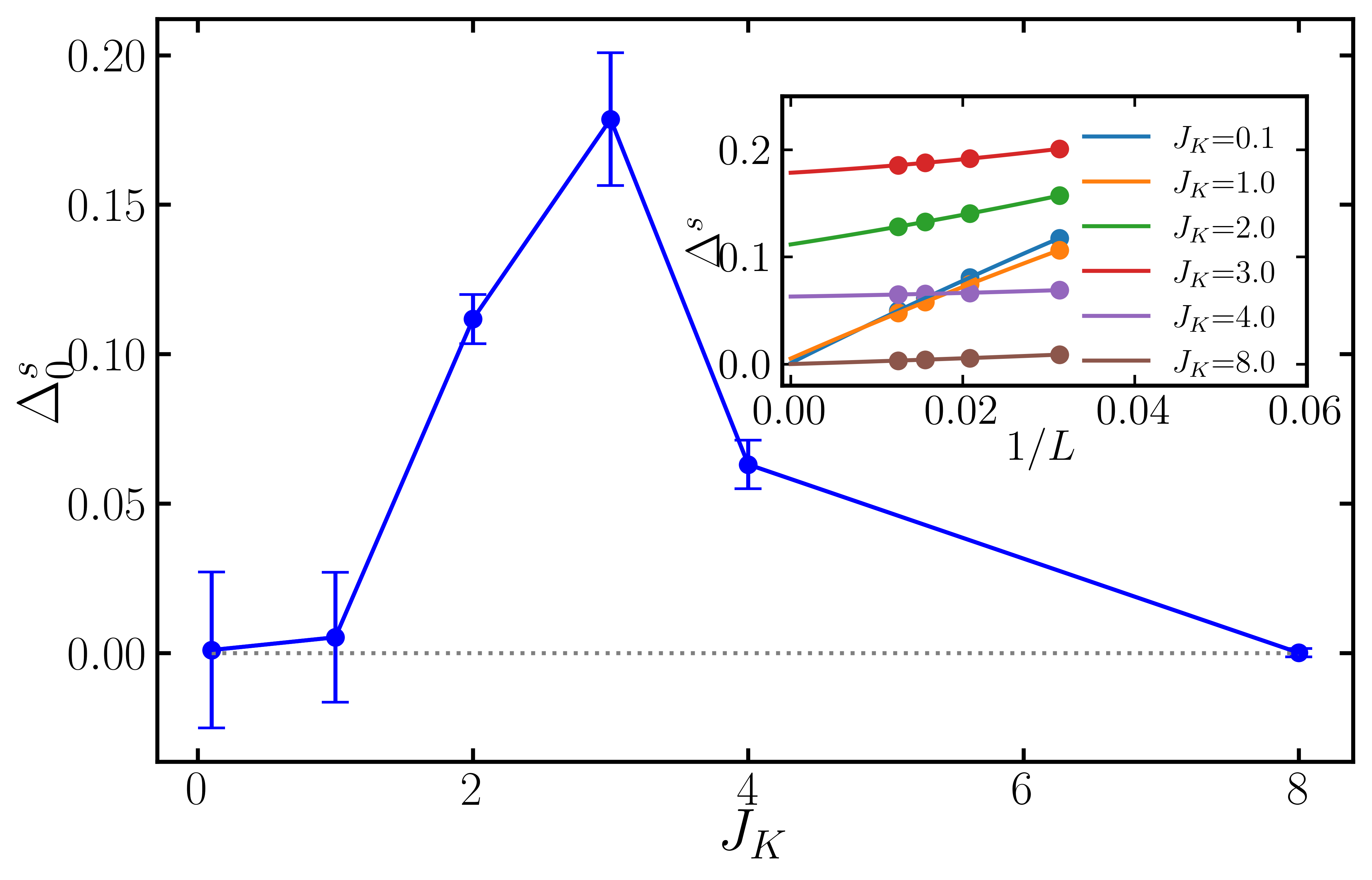}
	\caption{Spin gaps as a function of $J_K$ for $J_H=1$. Inset: spin gaps extrapolated to the thermodynamic limit for various values of $J_K$; the length of the chain ranges from $L=32$ to $L=80$.} \label{fig:spin_gap}
\end{figure}

\begin{figure}
	\centering
  \includegraphics[width=0.5\textwidth]{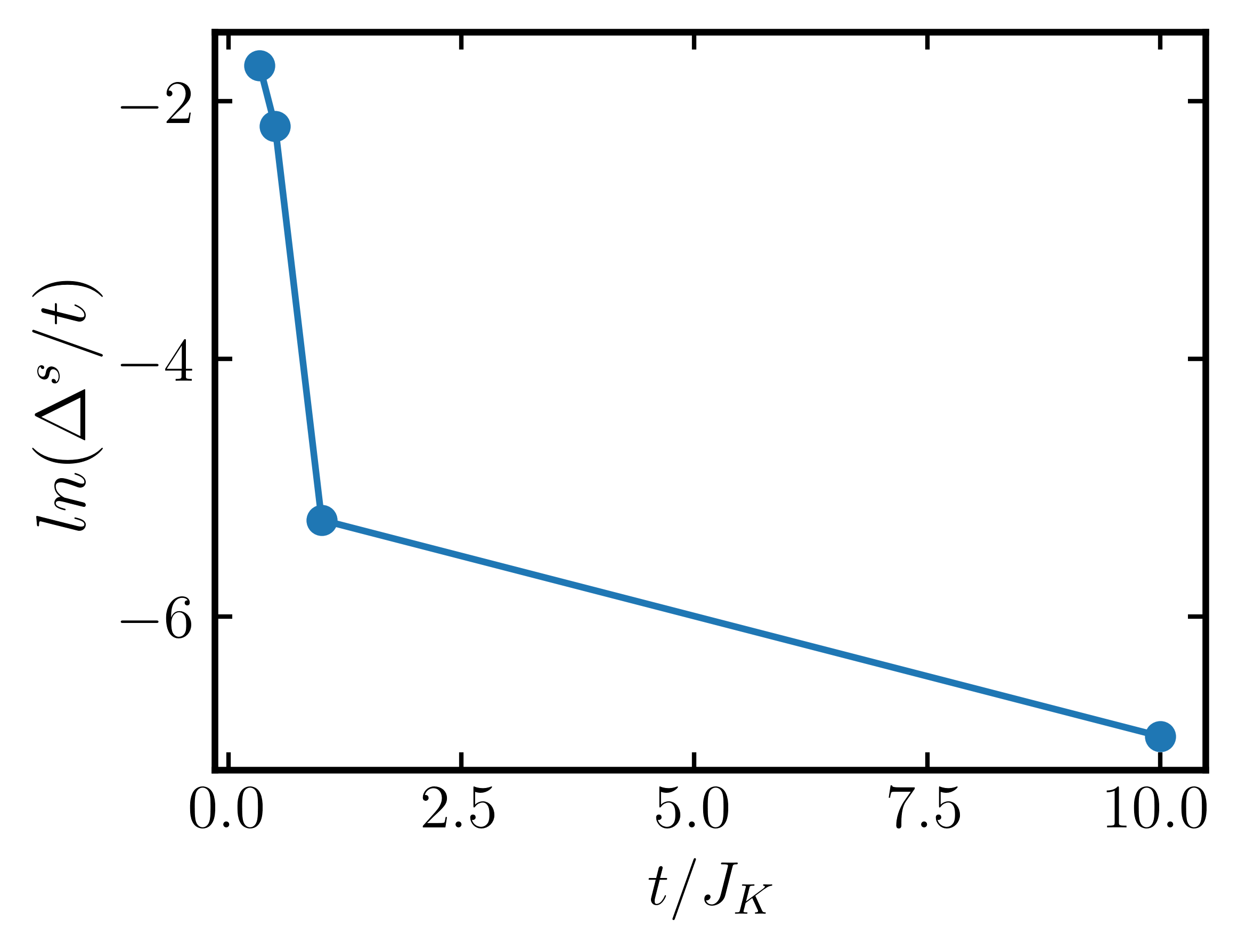}
	\caption{$ln(\Delta / t)$ vs. $t/J_K$ for $J_K=0.1$, 1, 2, and 3.} \label{fig:spin_bos}
\end{figure}

We present the spin gaps extrapolated to the thermodynamic limit in Fig.\ref{fig:spin_gap}. 
The spin gap is defined as:
\begin{equation}
    \Delta_s = E_0(N,S^z=1) - E_0(N,S^z=0)
\end{equation}
where $E_0(N,S^z)$ represents the ground state energy of a system with total electron number $N$ and spin $S^z$.
The spin gap has non-monotonic behavior with $J_K$, first increasing, and then decreasing to zero as the system transitions to the liquid phase at large $J_K$. In Fig.\ref{fig:spin_bos}, we show the relationship between $ln(\Delta /t)$ and $t/J_K$. In the bosonization study of Ref.\cite{Fujimoto1994,Sikkema1997}, these two quantities were predicted to obey a linear relationship. We re-examined this behavior over a wider range of $J_K$, and our results show that when $J_K$ is small ({\it i.e.}, $J_K<1$), the spin gap deviates from the bosonization prediction.

\begin{figure}
	\centering
  \includegraphics[width=0.7\textwidth]{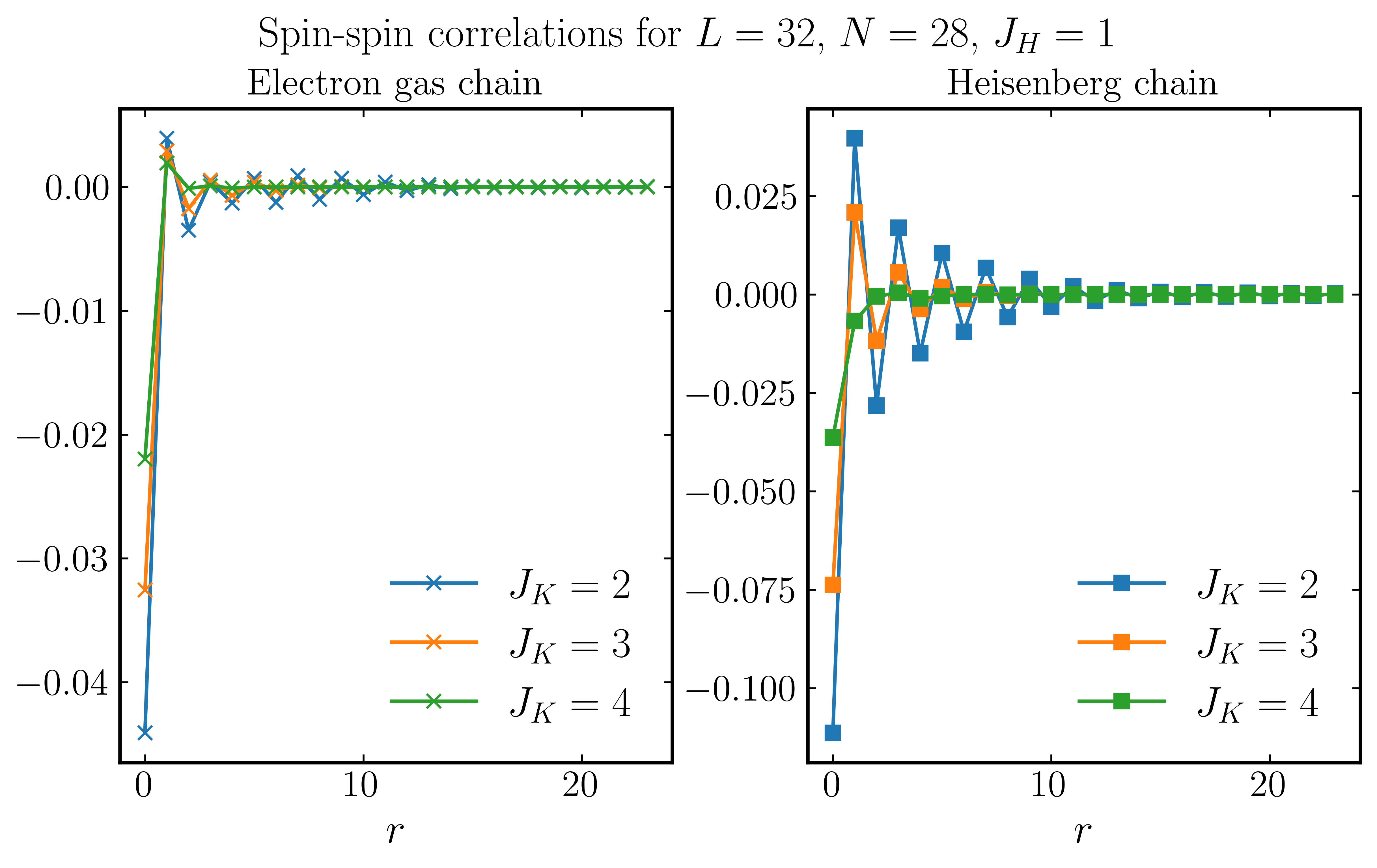}
	\caption{Spin-spin correlations measured on the electron gas chain (left) and Heisenberg chain (right).} \label{fig:spin_spin}
\end{figure}

In Fig.\ref{fig:spin_spin}, we show the spin-spin correlations on the electron gas chain (left) and the Heisenberg chain (right) for various values of $J_K$. We fix $J_H=1$ and the doping density at 1/8. The spin 
 correlation length keeps decreasing through the transition from the pair density wave (PDW) phase to the uniform superconducting (SC) phase. When the system enters the SC phase, only the nearest neighbors on the electron gas chain exhibit noticeable antiferromagnetic (AFM) correlations, which effectively vanish in the Heisenberg chain, even for nearest neighbors. The exponential decay of the spin-spin correlations in this spin-gapped phase can also be seen in Fig.\ref{fig:correlations} at $J_K=4$.

\section{Charge gaps extrapolated to the thermodynamic limit}
\label{sec:app_charge_gap}

\begin{figure}
	\centering
  \includegraphics[width=0.6\textwidth]{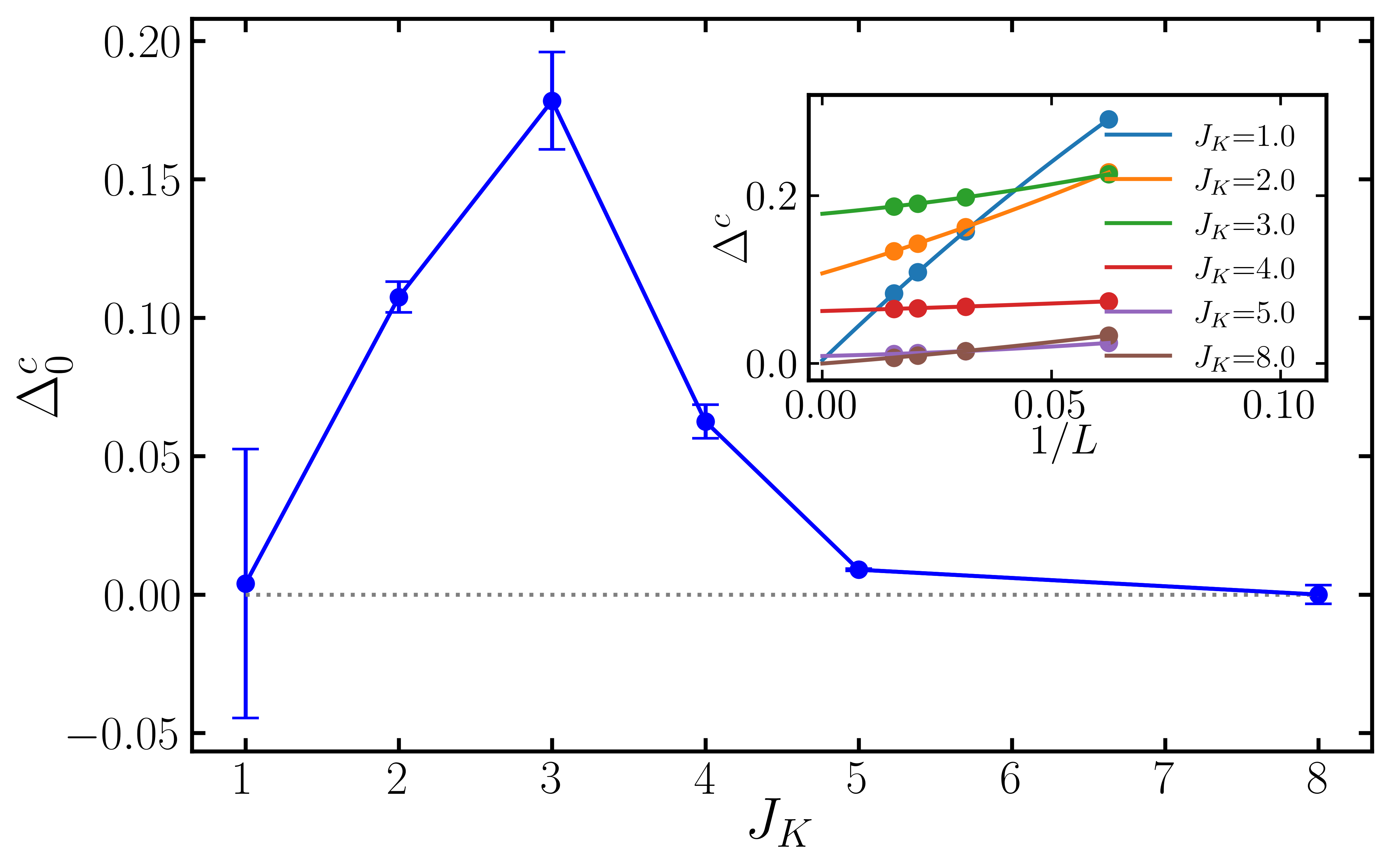}
	\caption{Charge gaps as a function of $J_K$ for $J_H=1$. Inset: charge gaps extrapolated to the thermodynamic limit for various values of $J_K$; the length of the chain ranges from $L=16$ to $L=64$.} \label{fig:charge_gap}
\end{figure}

The results for charge gaps extrapolated to the thermodynamic limit are displayed in Fig.\ref{fig:charge_gap}, in which the charge gap is defined as:
\begin{equation}
    \Delta_c = E_0(N+1,S^z=1/2) + E_0(N-1,S^z=1/2) - 2E_0(N,S^z=0)
\end{equation}
where $E_0(N,S^z)$ represents the ground state energy of a system with total electron number $N$ and spin $S^z$. Similar to the spin gap, the charge gap also exhibits a trend that first increases then decreases with the increasing of $J_K$. Furthermore, we observe from our numerical calculations that the charge gap and spin gap are identical for $J_K=2$, 3 and 4. The energy gap in this region corresponds to the energy that is required to break a bound pair, indicating that the system is in the pairing phase.

\begin{figure}[ht]
	\centering
  \includegraphics[width=0.4\textwidth]{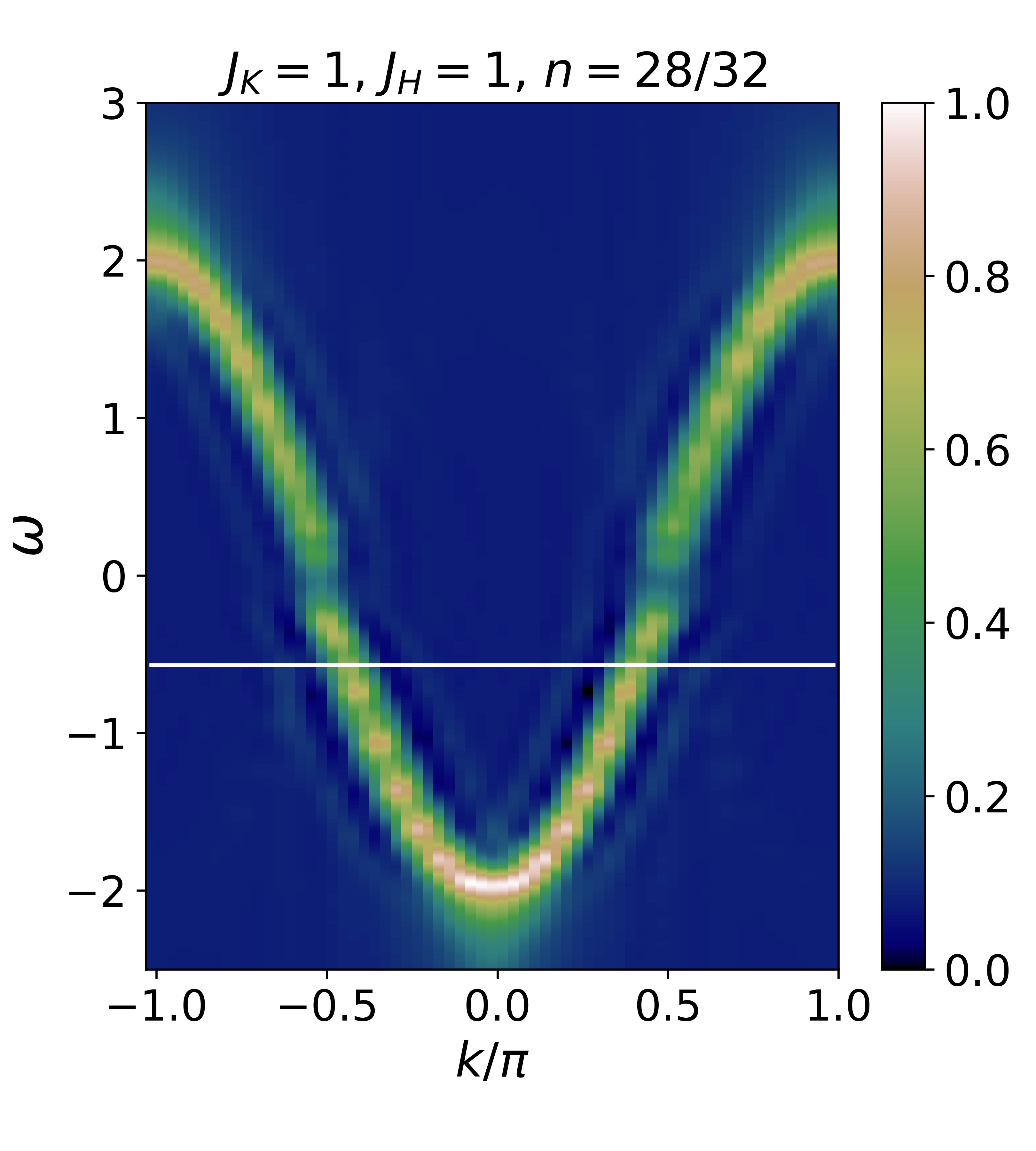}
  \includegraphics[width=0.4\textwidth]{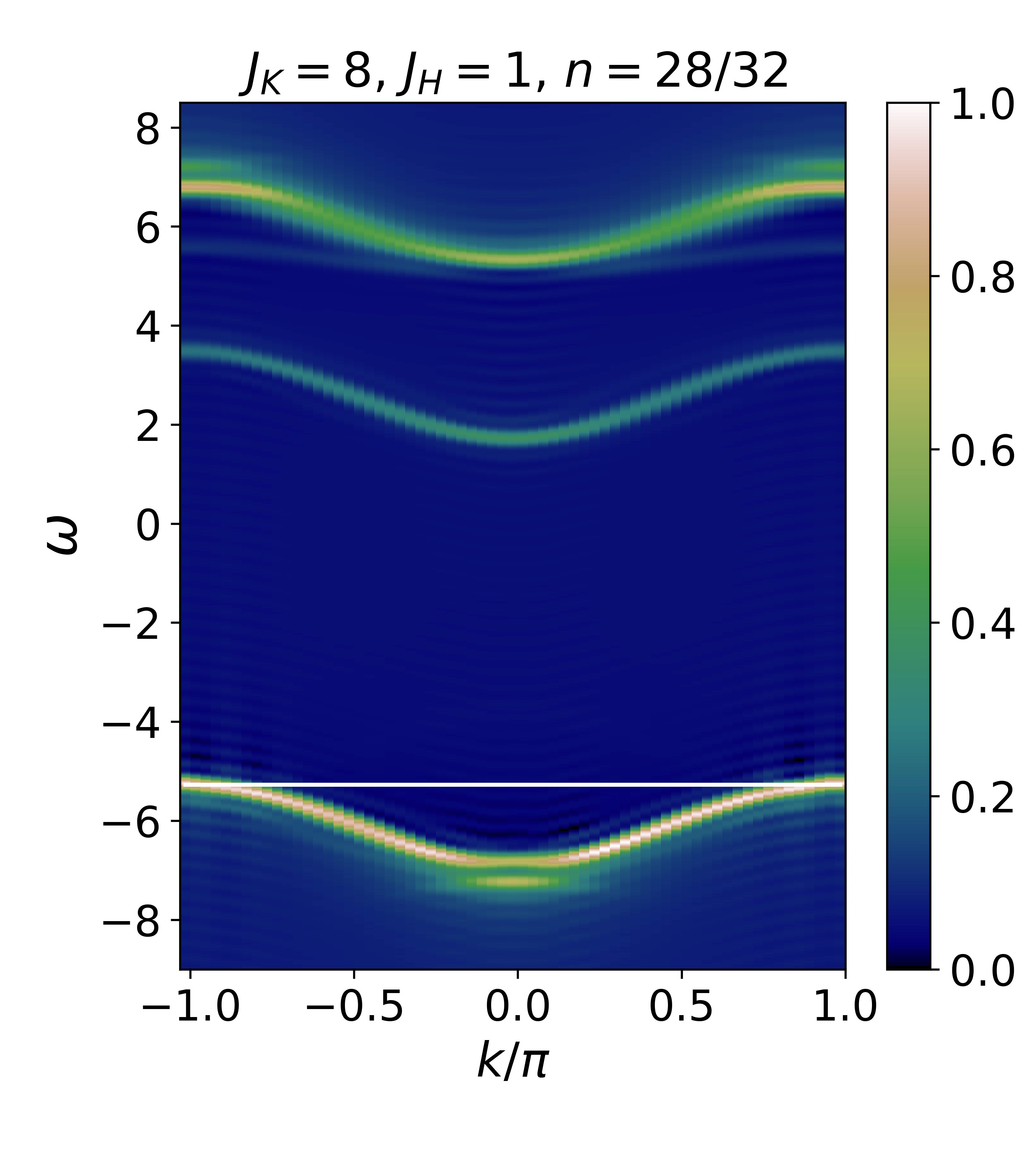}
	\caption{Normalized photoemission and inverse-photoemission spectra for $J_H=1$, $J_K=1$ (left) and $J_H=1$, $J_K=8$ (right). The single-particle-removal spectrum and single-particle-addition spectrum are separated by a thin white line denoting the Fermi energy.} \label{fig:photoemission_1}
\end{figure}

\section{Spectral functions in the small and large $J_K$ limits}
\label{sec:app_small_large_JK}

Photoemission and inverse-photoemission spectra for $J_K=1$ and $J_K=8$ are shown in Fig.\ref{fig:photoemission_1}, with white lines indicating the Fermi level. The spectrum for $J_K=1$ exhibits a small gap, although it is not clearly visible due to the computational resolution limitations. The spectrum for $J_K=8$ is gapless, and we observe quasi-coherent dispersive bands in this liquid phase. Although it was demonstrated that this is in a Luttinger liquid phase  in Ref.\cite{Sikkema1997}, we observe minimal signature of spin-charge separation.

\begin{figure}[ht]
	\centering
  \includegraphics[width=0.4\textwidth]{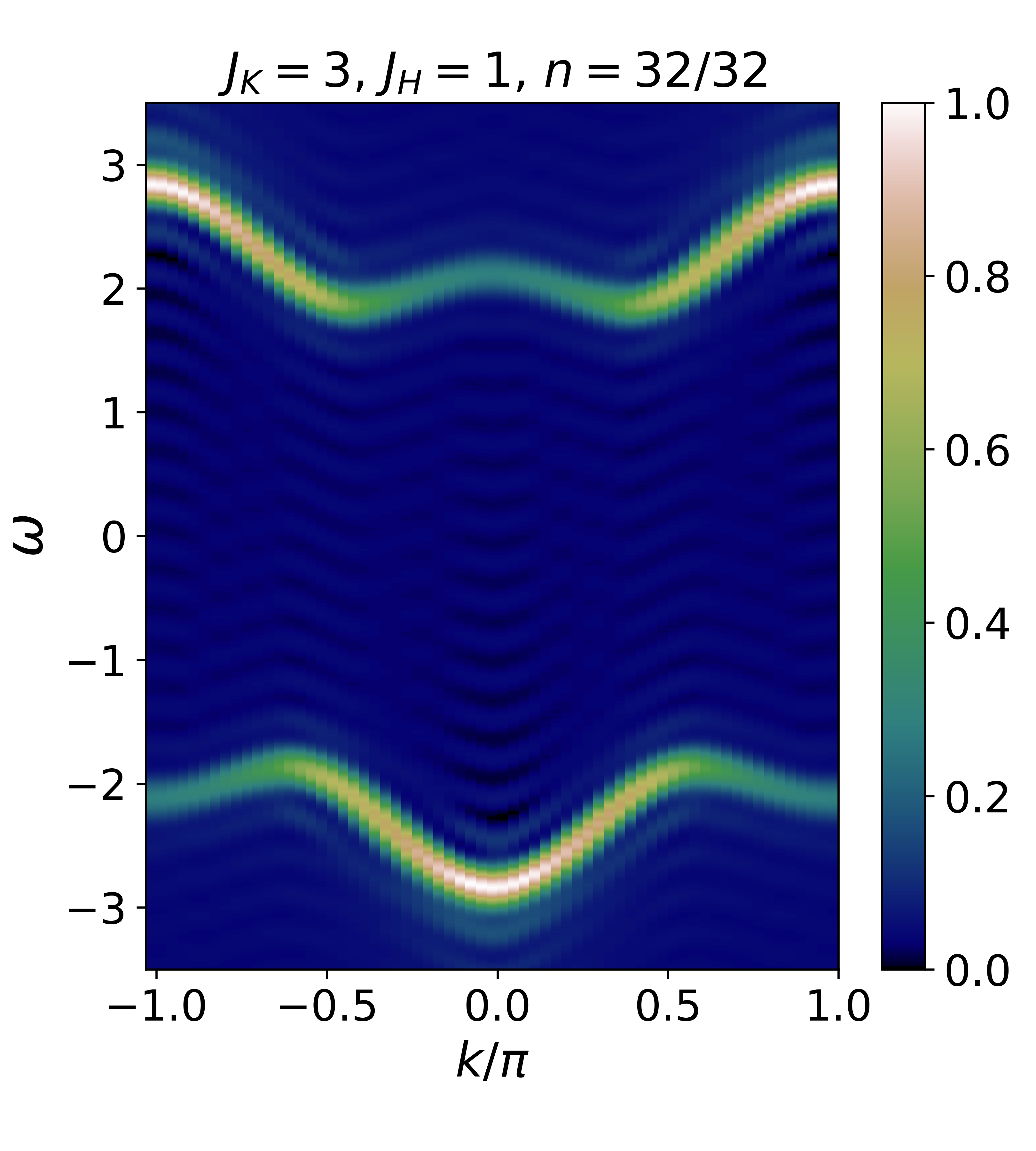}
	\caption{Normalized photoemission and inverse-photoemission spectra for $J_H=1$, $J_K=3$ at half filling.} \label{fig:photoemission_2}
\end{figure}

\begin{figure}
	\centering
  \includegraphics[width=0.4\textwidth]{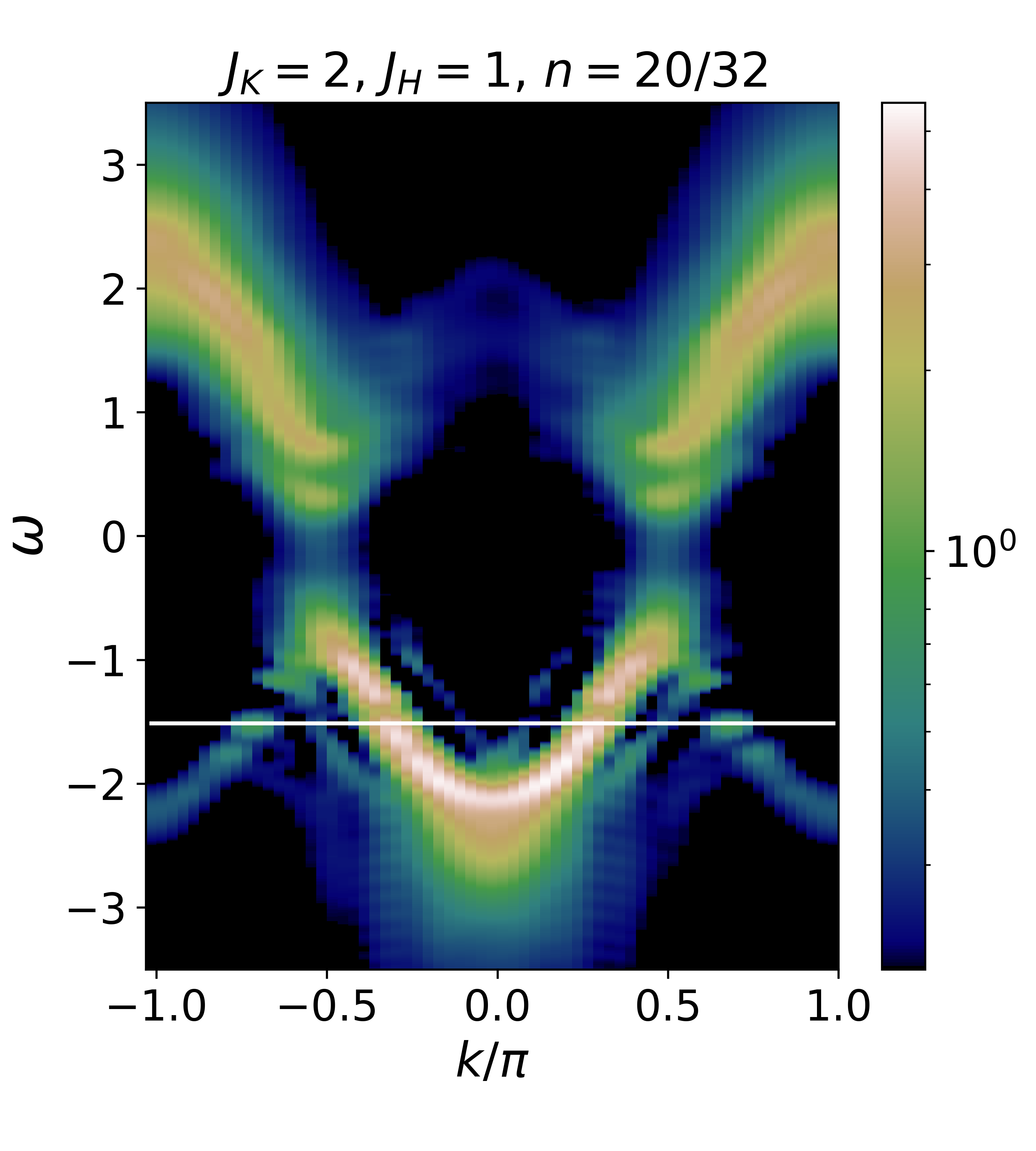}
  \includegraphics[width=0.4\textwidth]{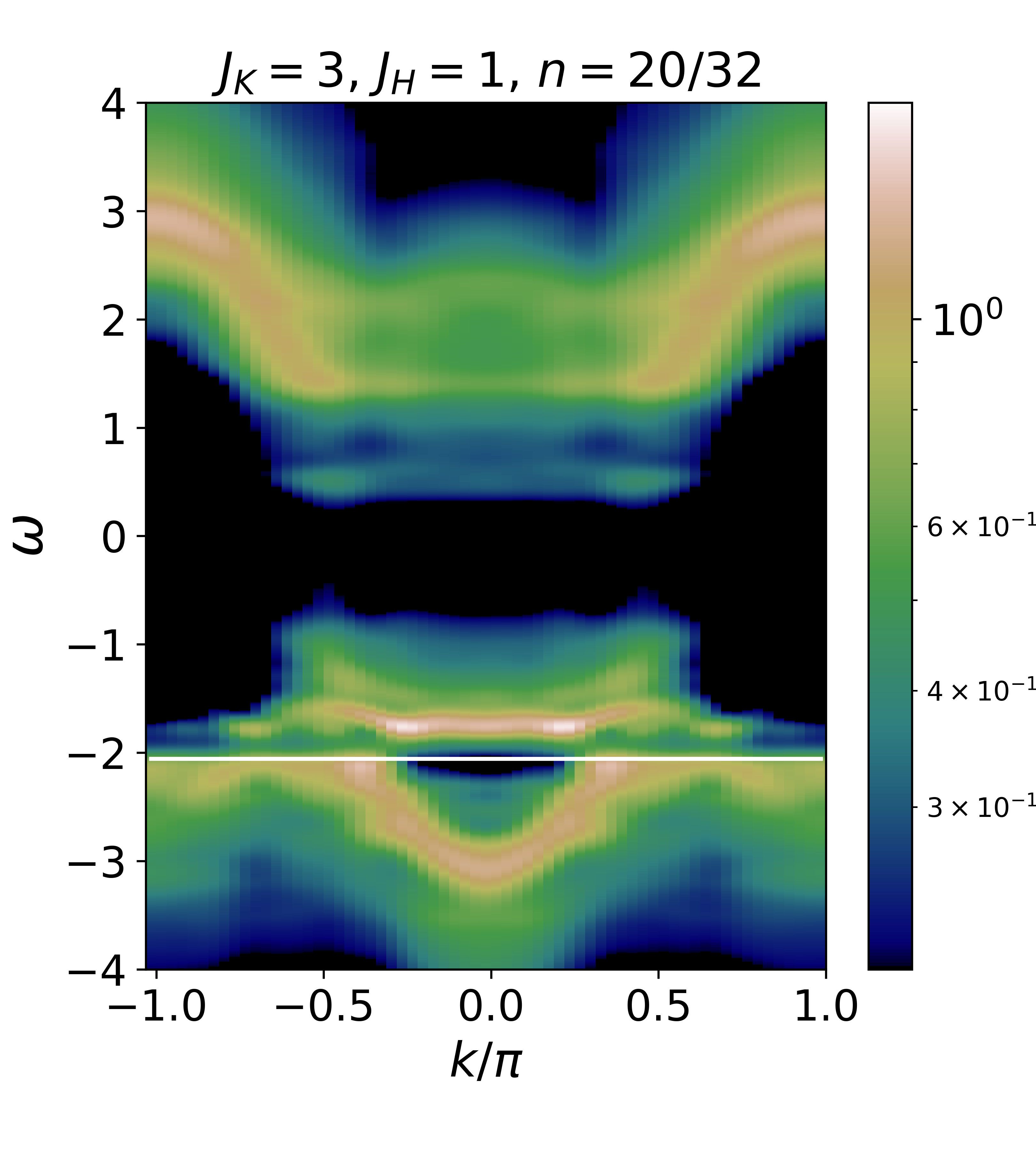}
	\caption{Photoemission and inverse-photoemission spectra for $J_H=1$, $J_K=2$ (left) and $J_H=1$, $J_K=3$ (right). The single-particle-removal spectrum and single-particle-addition spectrum are separated by a thin white line noting the Fermi energy. These results are presented in log scale colormaps.} \label{fig:photoemission_3}
\end{figure}

\section{Spectral functions at different filling densities}
\label{sec:app_filling}

The photoemission spectrum for $J_H=1$, $J_K=3$ at half-filling is presented in Fig.\ref{fig:photoemission_2}. The dispersion with two minima can be seen clearly in this result.

The single-particle removal and addition spectra for a higher doping density are shown in Fig.\ref{fig:photoemission_3}, where we present the data for $L=32$, $N=20$, $J_H=1$, and $J_K=2$ and 3. The four Fermi points are clearly visible in both cases.

\end{appendix}





\bibliography{ref.bib}


\end{document}